\documentclass[]{aa}
\usepackage{natbib}
\usepackage{graphicx}
\usepackage{wasysym}
\usepackage{txfonts}
\usepackage{color}
\def\WHI{\mbox{$\Upsilon_{{\rm HI}}$}}

\def\rmdv{\mbox{\rm dv}}
\def\fH2{\mbox{f$_\HH$}}
\def\fHI{\mbox{f$_{\rm HI}$}}

\def\EBV{\mbox{E(B-V)}}

\def\nH2{\mbox{${\rm n}_\HH}$}

\def\pccc{~{\rm cm}^{-3}} 
 
\def\pcc{~{\rm cm}^{-2}}

\def\Tsub#1 {\mbox{${\rm T}_{\rm #1}$}}
\def\TK  {\Tsub K }

\def\Tsp {\Tsub sp }

\def\p{\mbox{$^+$}}
\def\m1{\mbox{$^{-1}$}}

\def\h13cop{\mbox{{H$^{13}$CO\p}}}

\def\C3H{\mbox{C$_3$H}}
\def\c3h2{\mbox{C$_3$H$_2$}}
\def\cc3h2{\mbox{{\it c}-C$_3$H$_2$}}
 
 \def\R0{R$_0$}
\def\G0{\mbox{G$_0$}}

\def\ddeg{{}^\circ\kern-.1em}

\def \kms{\mbox{km\,s$^{-1}$}}

\def\E#1 {$10^{#1}$}
\def\E#1 {E{#1}}
\def\P#1,{$\nH2\TK~=~#1\times~10^4\pccc$~K}
\def\ec#1,#2,#3,{#1\,(#2)\E{#3}}

\def\H3{\mbox{H$_3$}}

\def\RH2{\mbox{R$_{\rm G}$}}
\def\g13{\mbox{g$_{13}$}} 

\def\cc3h{\mbox{{\it c}-\C3H}}
\def\lc3h{\mbox{{\it l}-\C3H}}

\newcommand{\emm}[1]{\ensuremath{#1}}   
\newcommand{\emr}[1]{\emm{\mathrm{#1}}} 

\newcommand{\hcop}{\emr{HCO^+}} 
 
\newcommand{\HH}{\emr{H_2}}

\def\t01{\tau_{0,1}}

\title{Comparing absorption from classic tracers of the diffuse interstellar medium 
at optical and radio wavelengths along the same sightlines}

\author{ H. Liszt\inst{1}}

\institute{
     National Radio Astronomy Observatory,
           520 Edgemont Road,
           Charlottesville, VA,
           USA 22903 
      \email{hliszt@nrao.edu}
}

\begin{document}
\date{received \today}%
\offprints{H. S. Liszt}%
\mail{hliszt@nrao.edu}%
%
\abstract
{}
{To compare information gained from radio and optical absorption line profiles 
from the diffuse interstellar medium along the same sightline}
{We compare new and existing $\lambda$21cm HI and $\lambda$3.4mm 
HCO$^+$ profiles with profiles of the optical tracers CaII, NaI, and KI
from an unpublished thesis of Tappe (2004)}
{The atoms traced optically are all heavily depleted compared to a Solar abundance and 
only the integrated optical depths of HI and HCO$^+$ correlate well with E(B-V).
HCO$^+$ is the species with by far the most limited kinematic distribution and the 
narrowest lines followed in order by KI, HI, NaI and CaII. CaII behaves separately in 
both column density and kinematics because it samples broader-lined warmer gas. Tracers 
of the cold neutral medium NaI, KI, HI and HCO$^+$ share the same kinematic space 
statistically without correlating to nearly the same extent in abundance. 
N(NaI) and N(KI) are correlated, as are the integrated optical depths of HI and 
HCO$^+$, but abundance correlations between optical and radio tracers are not seen. 
In the only direction with a measured CH$^+$ profile, a 2 km s$^{-1}$ velocity shift 
between CH$^+$ and CH, usually interpreted as the sign of shocked gas, is mimicked 
in the shift of HI relative to  HCO$^+$. CH and HCO$^+$ appear in the ratios 
N(CH)/N(HCO$^+$) = 14.6 and 21.1 along the two sightlines with optically-measured 
N(CH), compared with a mean of 12 determined previously at radio and submillimeter 
wavelengths. }
{}

\authorrunning{Liszt} \titlerunning{Optical and radio at the same time}

\maketitle{}

%

\def\lcm21{\mbox{{$\lambda$21cm}}}

\section{Introduction - An origin story}

Radio and optical astronomers study the diffuse interstellar medium in complementary 
ways but often talk past each other without recognizing the contributions of the 
other side of their discipline.  For instance the authoritative review 
article ``Diffuse Atomic and Molecular Clouds''  \citep{SnoMcC06} does not 
mention the \lcm21\ HI line of neutral atomic hydrogen 
\footnote{To be fair it also does not discuss CaII, NaI or KI}. 
The realization that polyatomic molecules had been studied in absorption in
diffuse clouds at radio wavelengths for some time \citep{CoxGue+88,LucLis93} 
was treated with disbelief in a preliminary draft of \citep{SnoMcC06}.

That said, \cite{SnoMcC06} did eventually come around, and they noted one work that 
explicitly bridged the gap, in which \cite{TapBla04} observed red-shifted atomic 
optical absorption lines against the optically
bright, radio-loud quasar B2145+067 whose Galactic \hcop\ absorption had previously 
been observed at mm-wavelengths by \cite{LucLis96}. \cite{SnoMcC06} noted a forthcoming
paper that never appeared, in which Tappe and Black were to discuss the  Galactic 
optical absorption toward B2145+067. Inspired by the idea that it might be possible
to compare profiles of \lcm21\ HI absorption with those of NaI and KI, 
we procured a copy of Tappe's unpublished 
thesis \citep{Tap04} on EBay and found that results were reported for five sightlines in 
sufficient detail that a full accounting of the observations and a reconstruction of 
the optical line profiles was possible.  In turn we completed the complementary roster
of \lcm21\ HI  and 89 GHz J=1-0 \hcop\ absorption spectra that allows the 
present discussion comparing a full panoply of Galactic optical and radio profiles, 
especially \lcm21\ H I, for the first time.

Section 2 summarizes the new and old observational material discussed here.  Section
3 is a detailed comparison of the optical (CaII, NaI, KI, CH, and CH\p) and radio
(HI and \hcop) line profiles toward the five sightlines noted in Table 1. Section 4 is 
a summary and section 5 is a coda giving some pertinent details not considered germain 
in the Introduction.
 
\section{Observations}

\subsection{Some conventions}

In this work, the rubric HI is used to denote neutral atomic hydrogen following
\cite{BohSav+78}.  N(H) is the column density of H-nuclei in atomic and 
molecular form, N(H) = N(H I)+2N(\HH)+N(H\p).  We denote the integrated optical depth 
of the $\lambda$21cm HI line as \WHI. Velocities presented with the spectra 
are measured with respect to the kinematic definition of the Local Standard of 
Rest. Optical reddenings \EBV\ are taken from the work of \cite{SchFin+98} and
these are converted to total hydrogen column densities as
N(H) $= 8.3\times 10^{21}$ \EBV\ $\pcc$ mag$^{-1}$ \citep{Lis14xEBV}. 

\begin{figure*}
\includegraphics[height=8.15cm]{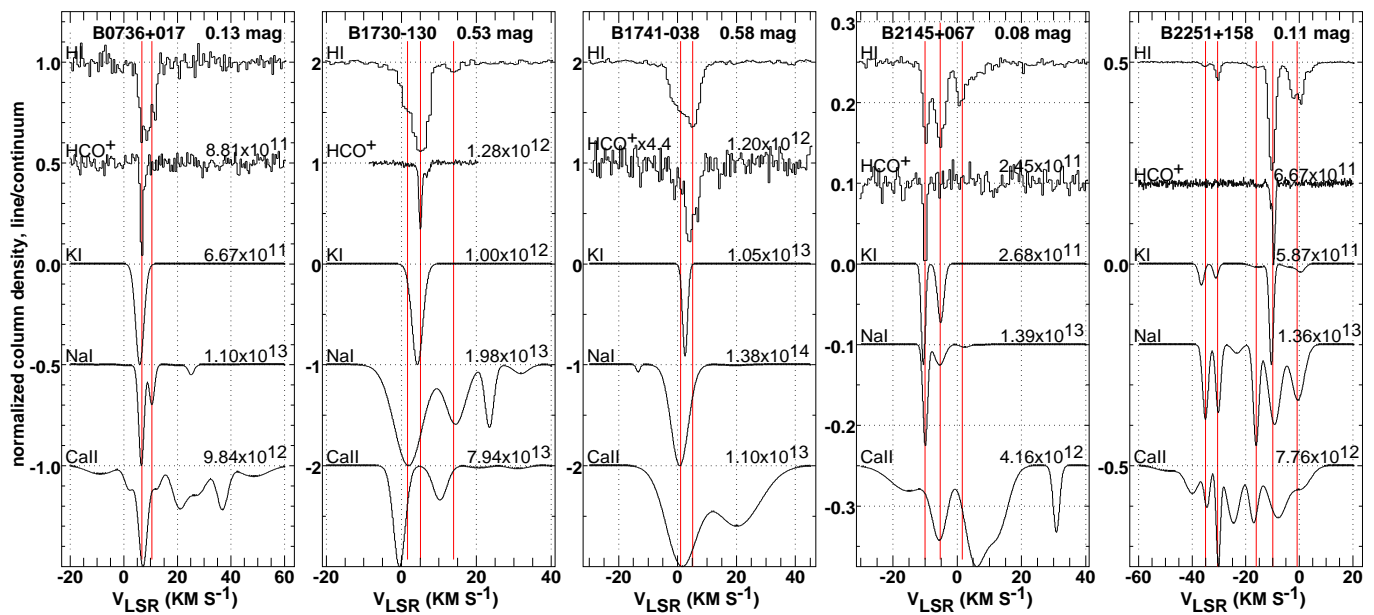}
  \caption[]{Spectra of the optical and radio tracers for all sources.  The HI and \hcop\ 
spectra at the top in each panel are observed line/continuum ratios.  Line profiles 
of the optical tracers KI, NaI and CaII are reconstructions of the gaussian decomposition 
of observed line profiles after deconvolution of the inherent atomic fine structure.  
The optical line  profiles are in units of column density, inverted and scaled to fit the 
vertical scale. Total column densities of \hcop, KI, NaI and CaII are shown with eath
profile. \EBV\ values from Table 1 are shown for each source. Weak absorption from CaII, 
and occasionally from NaI, extends beyond the velocity range shown here.}
\end{figure*}

\begin{table*}
\caption{Sightline properties and abundances$^1$}
{
\small
\begin{tabular}{lcccccccccc}
\hline
Target  & $l$  &  $b$     & \EBV$^2$  & N(CaII) &  N(NaI)       &  N(KI) &   N(CH$^+$) & N(CH) & N(\hcop) & \WHI   \\
      & deg  &  deg     &  mag  &  log $\pcc$ & log $\pcc$  & log $\pcc$  & log $\pcc $  & log $\pcc$& log $\pcc$ & \kms   \\
\hline
B0736+017 &  216.990 &   11.380 & 0.129 & 12.977 & 13.045 & 11.824 & 13.145 & 13.113 & 11.949 & 3.203 \\
      &       &       &       & (0.168) & (0.195) & (0.010) & (0.092) & (0.029) & (0.148) & (0.022) \\
B1730$-$130 &   12.032 &   10.812 & 0.526 & 13.900 & 13.296 & 12.000 &  &  & 12.106 & 10.927 \\
      &       &       &       & (0.200) & (0.197) & (0.098) &  &  & (0.030) & (0.088) \\
B1741$-$038 &   21.591 &   13.127 & 0.579 & 13.043 & 14.140 & 13.020 &  &  & 12.079 & 8.146 \\
      &       &       &       & (0.197) & (1.676) &   &  &  & (0.087) & (0.096) \\
B2145+067 &   63.656 &  -34.072 & 0.084 & 12.619 & 13.144 & 11.428 &  &  & 11.332 & 0.938 \\
      &       &       &       & (0.662) & (0.748) & (0.162) &  &  & (0.249) & (0.014) \\
B2251+158 &   86.111 &  -38.184 & 0.108 & 12.890 & 13.134 & 11.769 & $<$12.360 & 12.853 & 11.529 & 1.806 \\
      &       &       &       & (0.359) & (0.313) & (0.313) &  & (0.161) & (0.020) & (0.018) \\
\hline
mean &        &    & 0.285  & 13.35 & 13.59 & 12.41 &  &  & 11.89 & 5.00 \\
mean$^3$ &        &   & 0.212  & 13.37 & 13.16 & 11.79 &  &  & 11.83 & 4.22 \\
mean$^4$ &        &   & 0.118  & 12.94 & 13.09 & 11.80 &  &13.00  & 11.79 & 2.50 \\
\hline
\end{tabular}
\\
$^1$Quantities in parentheses are the standard deviation of the value immediately above
$^2$\cite{SchFin+98} $^3$Omitting B1741-038 $^4$For sources with measured N(CH) \\ 
}
\end{table*}

\subsection{Optical absorption measurements and line profiles}

\cite{Tap04} measured optical absorption spectra of CaII, NaI, and KI
toward the five compact extragalactic mm-wave continuum sources listed in 
Table 1 using the UVES instrument on the VLT, and observed CH and CH\p\
toward two sources, detecting CH\p\ in one case. The total column densities 
for these species given in Table 1 are taken directly from \cite{Tap04} who 
deconvolved the atomic fine structure
from his profiles and gave tables of Gaussian components fitting the deconvolved
profiles.  We reproduce Tappe's spectra in Figures 1 and 3 as the sum of these 
components rendered at 0.2 \kms\ intervals. The optical profiles are presented
 in units of column density and inverted for comparison with the radiofrequency 
HI and \hcop\ absorption spectra that are 
shown as observed line/continuum ratios.  The optical spectra are arbitrarily 
normalized to fit the vertical scale in Figures 1 and 3 but column densities are
shown with the spectra in Figure 1.  Sources with large column densities of
a species tend to have fewer apparent features in the optical profiles shown here
because  higher total column densities are typically produced by one component 
that is so much stronger than others that the weaker lines are scarcely visible.    

The NaI and KI column densities determined by \cite{Tap04} toward B1741-038 are 
discrepant (see Figure 2) and have very large or undetermined errors, respectively 
(Table 1). These values receive special treatment in Table 1 and as noted in the
discussion below when appropriate.

\subsection{Radiofrequency absorption measurements of HI}

$\lambda$21cm HI profiles toward B1730-130 and B1741-038 were taken by \cite{DicKul+83}.
The appearance of \cite{Tap04} spurred us to take HI profiles of the other sources at the 
eVLA as previously presented  in \cite{LisPet12} for all of the sources discussed here 
except B1741-038. \cite{LisPet12} also showed maps of CO J=1-0 emission
in sky fields around the background sources except B1741-038.

\subsection{Radiofrequency absorption measurements of \hcop}

\cite{LucLis96} observed \hcop\ toward all the sources discussed here, and improved 
profiles  were subsequently acquired at the PdBI instrument toward B1730-130 and 
B2251+158. Excitation temperatures are known to be very small in \hcop\ in 
diffuse molecular gas, given the weakness of emission that accompanies even 
very heavily saturated absorption lines \citep{LucLis96,Lis20Splotch}.  
We convert observed integrated \hcop\ 
optical depths to \hcop\ column density for a permanent dipole moment of 3.89 
Debye assuming rotational excitation in equilibrium with the cosmic microwave
background, ie N(\hcop) = $1.12\times 10^{12}\pcc \int \tau(\hcop)~\rmdv$.

\subsection{Abundances of CH and \hcop\ relative to \HH}

The abundances of CH and OH relative to \HH\ have been measured on many 
sightlines with the results that N(CH)/N(\HH) $= 3.6\pm0.8\times10^{-8}$ 
\citep{Lis07CO,SonWel+07,SheRog+08,WesGal+09,WesGal+10} and
 N(OH)/N(\HH) $= 1.0\pm0.3\times10^{-7}$ \citep{WesGal+09,WesGal+10}. In turn 
the N(\hcop)/N(OH) and N(\hcop)/N(CH) ratios have been measured at radio/submm 
wavelengths \citep{LisLuc96,GerLis+19} with the result that   
N(\hcop)/N(\HH) $= 3\times 10^{-9}$ with an estimated uncertainty of $\pm$0.2 dex.

The observations discussed here are the first cases where CH has been detected
in the optical waveband in directions with measured mm-wave \hcop\ absorption.
 
\section{Observational results}

General properties of the sightlines discussed here are given in Table 1,
along with the integrated optical depths \WHI\ of HI and total column densities 
of the other tracers.  Profiles for all sources are shown in Figure 1. 
Figure 2 plots total column densities and \WHI\ against \EBV, \WHI, N(NaI) 
and N(\hcop). 

\subsection{Abundances, depletions, correlations with \EBV\ {\it cum} N(H)}

Table 2 shows the percentages of a Solar abundance \citep{LodPal+09} 
that are represented by the total column densities of CaII, NaI and KI when 
N(H) $= 8.3\times 10^{21}$ \EBV\ $\pcc$ mag$^{-1}$ \citep{Lis14xEBV}. 
The fractional elemental abundances of sodium, calcium and potassium are 
small, at most $\approx 1$\%, and vary by factors of 7-8 among the various
sightlines. If the 
dominant ion stage is being observed in each case, the small, sharply 
varying fractional abundances must reflect strong and heavily variable 
elemental depletion along  moderately-reddened sightlines.


\subsection{Correlations between abundances}

The results shown in Figure 2 for the profile-integrated quantities are 
abstracted in a matrix of Pearson two-point r-function correlations in Table 3,
where values with at least $3\sigma$ significance are shown in bold. The
NaI and KI column densities determined toward B1741-038 were not included in this 
calculation owing to their aberrant behavior (Figure 2). For such a small sample 
only very high degrees of correlation have even formal statistical significance. 
Nonetheless the spectral line tracers seem quite tribal; NaI correlates only with 
KI, and \hcop\ with H I.

\EBV\ is a measure of the bulk column density and all tracers are expected to be
at least somewhat 
correlated with \EBV\ due to general mixing of conditions along the line of sight. 
The previously-demonstrated tight correlation of \WHI\ and \EBV\ \citep{Lis19} 
is strongly manifested even in this small sample. Given the known variability 
of the molecular gas fraction
\footnote{see for instance Figure 1 of \cite{LisPet+10}}, 
it is perhaps surprising that N(\hcop) is better correlated with \EBV\ than are 
any of the optical atomic gas tracers. Apparently, elemental depletion and ionization
equilibrium affect the optical atomic gas tracers to a greater extent than
HI and \hcop\ are affected by the influences on their detectability - 
the \HH\ fraction and ion molecule chemistry for N(\hcop), the kinetic temperature
and phase equilibrium for N(HI) and \WHI.  
  
The total abundances of NaI and KI are strongly coupled, despite some obvious
kinematic disparities in Figure 1, and uncorrelated with that of CaII. Given 
its wider kinematic distribution, CaII will sample the warm and intercloud 
gas to a greater extent than any of the other tracers, but CaII is also the 
optical species best correlated with \WHI\ that avoids warmer gas. 
This is presumably a bulk effect of a high degree of mixing in the ISM. 

Seeing the relationship of HI absorption to that of the canonical optical tracers 
was one of the most eagerly-anticipated results of this work. Despite the expectation
of a dominant contribution from the atomic cold neutral medium (CNM) to NaI and KI, 
it is CaII and \hcop\ whose column densities are best correlated with \WHI. This 
is somewhat in contradiction to the kinematics on display in Figure 1 as quantified next.

\subsection{Kinematics and kinematic correlations}

\hcop\ is the species with by far the most limited kinematic distribution and the 
narrowest lines in Figure 1, followed in order by KI, HI, NaI and CaII. To quantify 
the comparison of line profiles we interpolated H I, \hcop\ and the optical atomic 
tracers toward each source onto a common velocity grid and calculated a kinematic 
correlation coefficient r$_k$[m,n] between species m and n observed toward source k as

$$r_k[m,n] \equiv \int \hat{x}_m({\rm v}) \cdot \hat{x}_n({\rm v})~\rmdv 
\eqno{1}$$

where $\hat{x}_m({\rm v})$ and $\hat{x}_n({\rm v})$ 
are the mean-subtracted column density (for optical tracers) or optical depth 
profiles of tracers m and n toward source k, expressed as vectors with unit 
maxima, equivalent to the Pearson two-point correlation functions shown 
in Table 3.
This calculation used the profiles shown in Figure 1 and ignored some low column 
density features outside the velocity range shown, which appear mostly in CaII 
and less often in NaI.  Normalizing each profile mitigates the influence of 
correlations and uncertainties in abundance and on this account we included the
NaI and KI profiles toward B1741-038 in this calculation. Table 4 presents a matrix 
of the mean and standard deviation of this coefficient averaged over all directions 
k for the five tracers represented in Table 3.  Values above the $3\sigma$ 
level are again shown in bold.

A complete dearth of kinematic correlations above the $3\sigma$ level shows that
CaII occupies a different kinematic space from the other tracers, although only
narrowly for NaI. By contrast, the other tracers sampling the CNM and diffuse 
molecular gas all overlap significantly.

The KI profile is dominated by one strong component that overlaps with \hcop\ for 
the three sightlines having the highest molecular column densities (B0736, B1730, 
B1741). About one-third of the KI column density does not coincide with \hcop\ 
toward B2145 
and B2251, as manifested in several kinematic components that presumably 
represent the tail of high atomic abundance at small N(\HH) shown 
by \cite{WelHob01} and discussed in Sect. 3.5 here. The same behaviour 
is exaggerated in NaI except toward B2145.

\subsection{Optical CH and CH\p\ vs radio HI and \hcop}

Figure 3 plots spectra of CH, CH\p, \hcop\ and HI.  The displacement between
CH and CH\p\ toward B0736 that is sometimes interpreted as the sign of 
a slow shock \citep{EliWat80,MonFlo+88,FloPin98} is intriguingly mimicked
in HI.  The eccentricity of \hcop\ with respect to the HI profile toward
B2251 is typical \citep{LisLuc96,LisPet12} and CH is aligned with neither.

The N(CH)/N(\hcop) ratios observed toward B0736 and BN2251 are 14.6 and
21.1 $\pm 0.15$ dex, as compared with the mean ratio 12$\pm0.2$ dex implied 
by the discussion in Section 2.5 \citep{GerLis+19}.  This is reflected in 
slightly different implied moleculer hydrogen fractions from CH and \hcop\ 
in Table 3 but both CH and \hcop\ are consistent in indicating a higher 
molecular fraction toward B0736.

\subsection{How well do NaI or KI  trace H$_2$?}
 
\cite{WelHob01} showed that NaI and KI are well correlated with N(\HH) for
N(\HH) $\ga 10^{20} \pcc$, with substantial tails of appreciable atomic abundance 
at N(\HH) $\la 10^{19} \pcc$ and large scatter and a substantial fraction of
sightlines with very low N(NaI) and/or N(KI) at $10^{19} \pcc \la$
N(\HH) $\la 10^{20} \pcc$.  A column density N(\HH) $= 10^{20} \pcc$ corresponds 
approximately to N(\hcop) $= 3\times 10^{11} \pcc$ and only the line of
sight toward B2145+067 has such a small N(\hcop) in the cohort studied here.  
That being the case  it might be expected
that N(NaI) and N(KI) would correlate well with N(\hcop) but that is not the
case in Table 3: These species only correlate in velocity space (Table 4).  

In retrospect it seems that CH, not NaI or KI, should be used to trace 
\HH\ in the optical waveband as discussed in Sect. 2.5.

\begin{figure}
\includegraphics[height=9.45cm]{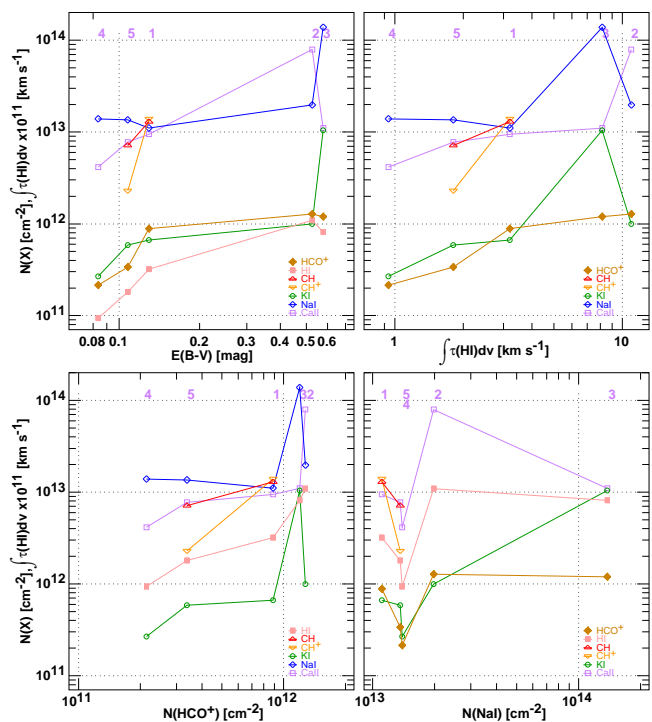}
\caption{Intercomparisons between \EBV, \WHI\ and measured column densities.  
\WHI\ has been scaled upward by $10^{11}$ to appear on the same scale. The numerals
at the top of each plot indicate which source's data is plotted at each abcissa,
following the right ascension order in Table 1 and the left-right order in Figure 1.}
\end{figure}

\begin{figure}
\includegraphics[height=11cm]{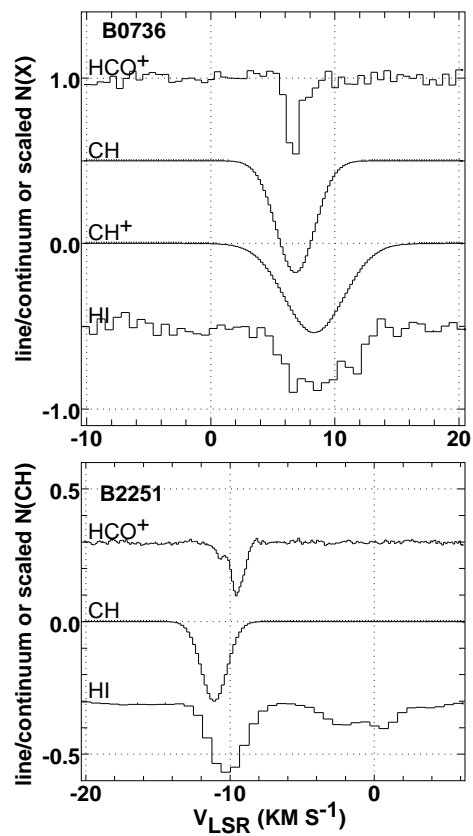}
\caption{Comparison of radio and optical molecular line profiles with HI. The vertical
scale is as in Figure 1.}
\end{figure}

\section{Summary}
 
Figure 1 shows the line profiles of the five species measured toward all five sources 
(Table 1): HI, \hcop, KI, NaI and CaII. HCO$^+$ has by far the most limited kinematic 
distribution and the narrowest lines followed in order in this regard by KI, HI, NaI 
and CaII. The atoms traced optically are all heavily depleted compared to a Solar 
abundance (Table 2).  The only abundance measures that correlate 
well with E(B-V) are the integrated optical depths of HI and \hcop\ (Table 3).

CaII behaves separately in both column density and kinematics because it samples 
broader-lined warmer gas. Tracers of the cold neutral medium (NaI, KI, HI and HCO$^+$) 
share the same kinematic space statistically with overlaps of 60-70\% (Table 4)
without correlating nearly as well in abundance (Table 3). N(NaI) and N(KI) are 
correlated, as are the integrated optical depths of HI and \hcop, but abundance 
correlations between optical and radio tracers are not seen in Table 3.

Four of the five sightlines have an NaI feature that is not matched in HI and only 
toward B2145 is there a feature (the highest velocity) in HI that is not matched 
in NaI. There are a few cases of HI features without KI counterparts in 
Figure 1, ie the 
highest velocity HI features toward B1730-130 and B2147+067 and perhaps the blue 
wings of the HI profiles toward B1730 and B1741, but no cases of KI features 
lacking a counterpart at some level in HI.
KI is notably more broadly distributed than \hcop\ toward B2145 and B2251, perhaps 
corresponding to a long-known tail of weak KI absorption at small N(H) and N(\HH) 
(see Section 3.5).
 
CH and \hcop\ appear in the ratios log N(CH)/N(\hcop) $= 1.2\pm0.15$ and 
1.3$\pm0.16$ along the two sightlines with measured N(CH) in Table 1, compared 
to an expected mean value of $1.1\pm0.2$ from other considerations discussed 
in Section 2.5, where the uncertainty considers only \hcop. In the only direction 
with a measured CH$^+$ profile, a 2 km s$^{-1}$ velocity shift between CH$^+$ and 
CH, usually interpreted as the sign of shocked gas, is intriguingly mimicked in 
the shift of HI relative to \hcop.


\section{Coda}

The origin story in the Introduction is a fabulist concoction \citep{Frankfurt2005}. 
I did not need to procure a copy of \cite{Tap04} on EBay because Achim Tappe and I 
were in touch when I provided some \hcop\ spectra and explained how to use them
while Achim was still a student.  Unfortunately, neither this experience nor my 
subsequent prodding of the authors of \cite{TapBla04} sufficed to elicit a paper 
describing the Galactic results. I subsequently completed the roster of HI and 
\hcop\ spectra, allowing the present bagatelle.

\begin{table}
\caption{Percentage fractional abundances}
{
\small
\begin{tabular}{lcccccc}
\hline

Source &CaII$^1$           & NaI$^1$           & KI$^1$             & CH$^2$    &\hcop$^3$  & HI$^4$   \\
       &     & \%Solar   &    & \fH2  & \fH2  & \fHI \\
\hline
B0736 &  0.41 &  0.52 &  0.47 & 67.16 & 55.03 & 32.63  \\
B1730 &  0.85 &  0.23 &  0.17 &  & 19.46 & 27.37 \\
B1741 &  0.11 &  1.44 &  1.65 & & 16.62 & 18.56 \\
B2145 &  0.28 &  1.01 &  0.29 &   & 20.58 & 14.78 \\
B2251 &  0.41 &  0.76 &  0.50 & 44.30 & 25.13 & 22.08 \\
\hline
\end{tabular}
\\
$^1$Using N(H)$= 8.3\times 10^{21}$ \EBV\ $\pcc$ mag$^{-1}$
and Solar abundances from \cite{LodPal+09}
$^2$2N(\HH)/N(H) if N(CH)/N(\HH) = $ 3.6\times 10^{-8}$  
$^3$2N(\HH)/N(H) if N(\hcop)/N(\HH) = $ 3.0\times 10^{-9}$  
$^4$N(HI)/N(H) when \Tsp\ = 60 K
}
\end{table}

\begin{table}
\caption{Quantity correlation coefficients$^1$}
{
\small
\begin{tabular}{lcccccc}
\hline
         & CaII    & NaI       & KI         & \hcop\    & HI          & \EBV  \\
\hline
 CaII   &         &  0.03     &  0.02      &   0.63     &   0.81     & 0.60    \\
        &         & (0.71)    & (0.71)      & (0.45)    & (0.34)     & (0.46)      \\
 NaI    &  0.03 &             & {\bf 0.99}  &   0.30    &   0.30     & 0.48    \\
        & (0.71) &            & (0.08)    & (0.67)      & (0.67)     & (0.62)    \\
 KI     &  0.02 & {\bf 0.99} &           &   0.35       &   0.31     & 0.48   \\
        & (0.71) & (0.08)     &           & (0.66)      & (0.67)     & (0.62)    \\
\hcop   &   0.63 &   0.30     &  0.35     &             & {\bf 0.92} & {\bf 0.88}    \\
        & (0.45) & (0.67)     & (0.66)    &            & (0.22)     & (0.28)         \\
 HI     &   0.81 &   0.30     &  0.31     & {\bf 0.92}  &            & {\bf 0.95}   \\
        & (0.34) & (0.67)     & (0.67)    &  (0.22)    &            & (0.19)    \\
 \EBV   & 0.60   &   0.48     &  0.48     & {\bf 0.88} & {\bf 0.95} &          \\
        & (0.46) & (0.62)     & (0.62)    & (0.28)     & (0.19)     &        \\
\hline
\end{tabular}
\\
$^1$ Quantities in parentheses are the formal standard deviation of
the value immediately above \\
}
\end{table}

\begin{table}
\caption{Mean kinematic correlation coefficients$^1$}
{
\small
\begin{tabular}{lccccc}
\hline
      &  CaII       & NaI        &       KI   &    \hcop   &    HI\\
\hline
 CaII &           &0.58    & 0.32        & 0.29        & 0.47  \\
      &            & (0.22)   & (0.22)    & (0.25)       & (0.21)  \\
  NaI &  0.58 &        &   {\bf 0.65}  & {\bf 0.58}         &{\bf  0.71}  \\
      & (0.22)   &          & (0.14)    & (0.19)       & (0.08)  \\
   KI & 0.32    & {\bf 0.65} &           &  {\bf 0.58}        & {\bf 0.72}  \\
      & (0.22)   & (0.14) &              & (0.06)      & (0.15)  \\
\hcop & 0.29 &   {\bf 0.58} & {\bf 0.58} &             & {\bf 0.66}  \\
      & (0.25) & (0.19) & (0.06)       &             & (0.16)  \\
   HI & 0.47 & {\bf 0.71} & {\bf 0.72} & {\bf 0.66} &         \\
      & (0.21) & (0.08) & (0.15)    & (0.16) &         \\
\hline
\end{tabular}
\\
$^1$Quantities in this table are the mean over all five sources  \\ 
of the kinematic correlation coefficient defined in Eqn 1 and \\
quantities in  parentheses are the standard deviation of that mean \\
}
\end{table}

\begin{acknowledgements}

The National Radio Astronomy Observatory is a facility of the National Science 
Foundation operated under cooperative agreement by Associated Universities, Inc.
I thank Evelyne Roueff for discussions of the early draft text of \cite{SnoMcC06} 
at the Asilomar astrochemistry meeting in 2005.

\end{acknowledgements}

\bibliographystyle{apj}

\end{document}